\documentclass[runningheads]{llncs}

\usepackage[T1]{fontenc}
\usepackage{graphicx}
\usepackage{hyperref}
\usepackage{color}
\usepackage[colorinlistoftodos]{todonotes}
\usepackage{amsmath}
\usepackage{booktabs}

\begin{document}

\title{Towards Effective Human Performance \\in XR Space Framework based on \\Real-time Eye Tracking Biofeedback}

\titlerunning{Towards XR Framework for developing adaptive immersive systems}

\author{Barbara Karpowicz\inst{1,2}\orcidID{0000-0002-7478-7374} \and 
Tomasz Kowalewski \inst{1,2}\orcidID{0009-0002-0542-9546}\and
Pavlo Zinevych \inst{1,2}\orcidID{0009-0008-9250-8712}\and
Adam Kuzdraliński\inst{1}\orcidID{0000-0003-2383-1950} \and
Grzegorz Marcin Wójcik \inst{3}\orcidID{0000-0003-2820-397X}\and
Wiesław Kopeć\inst{1,2}\orcidID{0000-0001-9132-4171}
}

\authorrunning{B. Karpowicz et al.}
\institute{XR Center, Polish-Japanese Academy of Information Technology
\email{karpowicz.b@pja.edu.pl, kopec@pja.edu.pl}\\
\url{https://www.xrc.pja.edu.pl}
\and XR Space \url{https://xrspace.xrlab.pl/}
\and
Maria Curie-Skłodowska University in Lublin
}

\maketitle              
\begin{abstract}
This paper proposes an eye tracking module for the XR Space Framework aimed at enhancing human performance in XR-based applications, specifically in training, screening, and teleoperation. This framework provides a methodology and components that streamline the development of adaptive real-time virtual immersive systems. It contains multimodal measurements - declarative in the form of in-VR questionnaires and objective, including eye tracking, body movement, and psychophysiological data (e.g., ECG, GSR, PPG). A key focus of this paper is the integration of real-time eye tracking data into XR environments to facilitate a biofeedback loop, providing insight into user attention, cognitive load, and engagement. Given the relatively high measurement frequency of eye tracking - recognized as a noninvasive yet robust psychophysiological measure - this technology is particularly well suited for real-time adjustments in task difficulty and feedback to enhance learning and operational effectiveness. Despite its established role in cognitive and attentional studies, implementing eye tracking metrics within dynamic, real-time XR environments poses unique challenges, particularly given the complex moving visuals presented in head-mounted displays (HMDs). This paper addresses these challenges by focusing on the essential aspects of integrating eye tracking in immersive systems based on real-time engines, ultimately facilitating more efficient, adaptive XR applications.

 \and Human Performance \and Psychophysiology \and Eye tracking 
\end{abstract}

\section{Introduction}
Virtual Reality (VR) and eXtended Reality (XR) technologies are transforming the fields of training, screening, and teleoperation by enabling the creation of immersive virtual environments that preserve ecological validity. When VR or XR technology is used, a high degree of immersion can be achieved, making participants feel as if they are in a real environment. This increases the naturalness of their reactions and behavior. \cite{kisker2021,schudy2023} 
The aspect of preserving ecological validity conditions in training or screening is important as it can affect the realism of user responses, the transferability of learned skills to real-world situations, and overall engagement and effectiveness of the XR experience. XR-based systems enable realistic scenarios and interactions that may be difficult, expensive, dangerous, ethically controversial or even impossible to achieve in real life situations. \cite{mcintosh2022dialing} XR Space Framework is intended for precise control over training variables and conditions, allowing for consistent and repeatable scenarios that enhance skill acquisition and assessment accuracy. This approach combined with multimodal measurements - declarative in the form of questionnaires in virtual reality, without the need to remove the VR goggles, and objective, including body movement, eye tracking to measure executive/cognitive function, and psychophysiological data (e.g., ECG, GSR, PPG) - allows for comprehensive insights into user states without disrupting immersion. In particular, eye tracking operates at relatively high frequencies, enabling \textbf{Dynamic Difficulty Adjustment (DDA)}, feedback, and support based on the user's cognitive load. By continuously monitoring user responses, the XR Space Framework makes real-time adjustments possible, creating a more responsive and effective environment for learning and operational tasks.

The main goal of this paper is to present an integration of real-time eye tracking data with virtual immersive systems to \textbf{enhance human performance}. Eye tracking technology provides invaluable insights into user attention, cognitive load, and engagement by monitoring gaze patterns (based on \textbf{saccades}), \textbf{fixations}, and \textbf{pupil responses}. By incorporating these metrics, the framework enables the design of systems that dynamically adjust task difficulty, feedback, and learning pathways based on individual user performance in real-time.

\section{Related Works}
\begin{sloppypar}
Based on our previous research on effective immersive systems for multimodal data acquisition, we have successfully implemented real-time acquisition of eye tracking, body movement, and psychophysiological data across various XR-continuum immersive applications. These systems have been applied in numerous psychology research studies, yielding valuable insights into user attention, cognitive load, and engagement in immersive environments.\cite{karpo2022,kopec2023human,pochwatko2023invisible,pochwatko2023wellbeing,schudy2023} While the data acquisition phase is well-established, current efforts focus on developing pre-processing and processing pipelines that enable real-time analysis and adaptive responses. This includes dynamically adjusting task difficulty and feedback based on the user’s cognitive and physiological state, which remains a critical step in enhancing the effectiveness and personalization of XR-based training and screening systems.
\end{sloppypar}

\subsection{Data Aquisition}
\label{sec:dataAquisition}
The data acquisition process starts with \textbf{calibration}. In this state, the user is guided through a procedure to ensure accurate eye tracking performance. This typically involves adjusting the position of HMD and IPD (interpupillary distance), and lastly, the user looking at specific points in the virtual environment, allowing the system to align gaze data effectively. In our previous systems, 5-point calibration was used. \cite{karpo2022,kopec2023human,pochwatko2023invisible,pochwatko2023wellbeing,schudy2023}

\textbf{Real-time acquisition} of eye tracking data occurs continuously during the immersive experience. The system logs key metrics, such as gaze direction, pupil size, and blink status, using the integrated eye tracking SDK.

To manage the incoming data efficiently, a \textbf{buffering system} is implemented. This temporary storage holds eye tracking data points for a brief period, ensuring a smooth saving process and preventing data loss during high-frequency sampling. The collected data are saved to a file in a separate thread. This logged data can be used for a later analysis, enabling deeper insights into user behavior and cognitive load.

\subsection{Dynamic Difficulty Adjustment}
Research on the topic of human motivation, performance, and emotions, based on the Flow Theory \cite{csikszentmihalyi1992flow} has been already integrated with the field of Human-Computer Interaction (HCI), and led to the emergence of the study area named Affective Computing (AC) \cite{rosalind1997,rodrigues2023}. 

\begin{figure}
    \centering
    \includegraphics[width=1\linewidth]{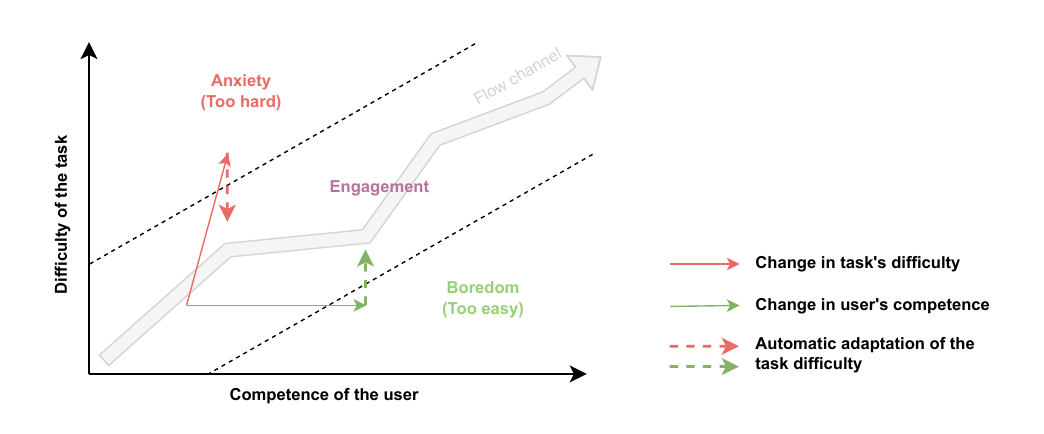}
    \caption{Flow chart with suggested automatic adjustment of task difficulty (source: own elaboration, based on \cite{csikszentmihalyi1992flow})}
    \label{fig:flowDiagram}
\end{figure}

When performing a task, the change between two emotional states can occur due to one of the following reasons\cite{chanel2011}:
\begin{itemize}
    \item User's competence increases, but task difficulty stays the same - can cause boredom (green solid arrow in Figure \ref{fig:flowDiagram})
    \item Task's difficulty increases too fast in comparison to user's competence increase - can cause anxiety (red solid arrow in Figure \ref{fig:flowDiagram})
\end{itemize}

The challenge lies in adapting task difficulty based on human performance. One approach is implementing Dynamic Difficulty Adjustment using eye tracking data - specifically saccades, fixations, and pupil responses. This paper aims to provide a proposition solution for the \textbf{pre-processing and processing of eye tracking data in a real-time graphic engine}, which can subsequently be applied to DDA. 

\begin{table}[]
\caption{Examples of eye tracking metrics for assessing engagement and emotional state}
\label{tab:eyeMetrics}
\begin{tabular}{p{0.15\linewidth} | p{0.3\linewidth} | p{0.5\linewidth}}
\hline
\multicolumn{1}{|c|}{\textbf{Metric}}  & \multicolumn{1}{c|}{\textbf{What it measures}}                 &  \multicolumn{1}{c|}{\textbf{\begin{tabular}[c]{@{}c@{}}Importance in engagement\\ /emotional state\end{tabular}}}                   \\ \hline
\begin{tabular}[c]{@{}l@{}}Fixation\\ duration\end{tabular}             & Length of focus on a point                            & Long fixations indicate high engagement, short fixations suggest scanning or disinterest.                                           \\ \hline
\begin{tabular}[c]{@{}l@{}}Saccade\\ velocity\end{tabular}                & Speed of gaze movement between points                 & High velocity indicates excitement or stress, low velocity may suggest calm focus.                                                  \\ \hline
\begin{tabular}[c]{@{}l@{}l@{}}Pupil\\ dilation\\
/constriction
\end{tabular}       & Changes in pupil size                                 & Larger pupils indicate higher cognitive load or emotional arousal. Small pupils may indicate relaxation, boredom, or disengagement. \\ \hline
Blink rate                    & Frequency of blinks                                   & Frequent blinking can signal disengagement, while fewer blinks suggest intense focus.                                               \\ \hline
\begin{tabular}[c]{@{}l@{}}Time to first\\ fixation\\
(TTFF)
\end{tabular}
  & Time to focus on a key element after stimulus         & Short TTFF indicates recognition or interest. Long TTFF suggests confusion.                                                         \\ \hline
Dwell time                    & Total time spent looking at a specific object or area & Long dwell time on relevant items indicates engagement. Short dwell time shows distraction.                                         \\ \hline
\end{tabular}
\end{table}

\section{Universal Solution Proposal}
Based on our previous research and the above mentioned research context, as well as state-of-the-art solutions, including HMDs with ET and contemporary RTCG engines, we propose the following universal approach toward universal implementation of DDA, based on specific ET metrics and theoretical background. 
\subsection{Eye Tracking Metrics}
\textbf{Eye tracking metrics} offer valuable insights into user engagement, involvement, and emotional state by capturing the unconscious and subtle movements of the eyes.

In table \ref{tab:eyeMetrics} there are examples of eye tracking metrics, implemented in the proposed module, that can be used to assess engagement and emotional state.

By monitoring gaze patterns and pupil responses, it is possible to provide adaptive, more personalized feedback and instructions during training or screening sessions. Additionally, it gives a possibility to \textbf{evaluate human performance}. If the user's gaze aligns well with tasks objectives (e.g., looking at key objects or following expected patterns), the level of difficulty can be increased. If the system detects missed fixations or gaze moving too quickly over a crucial area, it gives information that this person needs more guidance and further training first, before increasing the level of difficulty. 

\subsection{Theoretical Background}
\textbf{Velocity-Threshold Identification} is one of the most popular algorithms for detecting saccades and fixations based on eye tracking data. It is based on measuring the angular velocity of eye movement and comparing it with a set threshold to determine when there is fixation and when there is saccade. \cite{duchowski2022,Imaoka2020} \textbf{Eye movement angular velocity} is calculated from the difference of gaze positions over successive samples time (see Equation \ref{eq:angularVelocity}). 

\begin{equation}
\label{eq:angularVelocity}
    v=\frac{\theta}{\Delta t}
\end{equation}

where:
\begin{itemize}
    \item $\theta$ - angle between gaze direction vectors at moment of $t_{1}$ and $t_{2}$ 
    \item $\Delta t$ - time difference between $t_{1}$ and $t_{2}$
\end{itemize}

Equation \ref{eq:theta} presents the calculation of an \textbf{angle ($\theta$)} between the gaze direction vectors in successive samples time.

\begin{equation}
\label{eq:theta}
    \theta = arccos(\frac{P1 \cdot P2}{\left| P1 \right|\left| P2 \right|})
\end{equation}

where:
\begin{itemize}
    \item $P1$ - gaze direction vector at the moment of $t_{1}$
    \item $P2$ - gaze direction vector at the moment of $t_{2}$
    \item $P1 \cdot P2$ - scalar product of $P1$ and $P2$
    \item $\left| P1 \right|\left| P2 \right|$ - length of $P1$ and $P2$
\end{itemize}

Due to the fact that vectors $P1$ and $P2$ are normalized\footnote{Normalize vector has a length 1}, the denominator of this equation can be omitted, resulting in the following equation \ref{eq:theta2}: 

\begin{equation}
    \theta = arccos(P1 \cdot P2)
    \label{eq:theta2}
\end{equation}

The decision whether a given sample is part of a saccade of fixation is based on a given threshold.

\[v > v_{threshold1} \rightarrow saccade\]
\[v < v_{threshold2} \rightarrow fixation\]

Typically saccades are roughly bounded by 250 deg/s and fixations can pottentially be identified with angular velocity less than 3 deg/s. \cite{duchowski2022}

\subsection{Universal Approach}
The proposed solution utilizes the SRanipal API (associated with the HTC Vive Pro Eye headset). However, the module was created with a universal approach. The core advantage of this module lies in its flexibility and extensibility. It can integrate with a range of eye tracking hardware and software platforms, such as Varjo\footnote{\url{https://developer.varjo.com/docs/native/varjo-native-sdk}}, Unity OpenXR\footnote{\url{https://docs.unity3d.com/Packages/com.unity.xr.openxr@1.13/manual/features/eyegazeinteraction.html}}, and others. When it comes to Unity OpenXR, its eye tracking input data is limited only to eye position and rotation (lacking data needed for pupillometry). Therefore, it can be used only for saccades and fixations, without pupil responses. 

The universal approach is achieved by implementing abstract classes and interfaces, which act as standard templates for processing eye tracking data. Developers can easily extend these abstract classes to accommodate new data sources, ensuring that the module can adapt to future technologies or different platforms without significant rework.

\subsection{Data Workflow}
\begin{figure}
    \centering
    \includegraphics[width=0.7\linewidth]{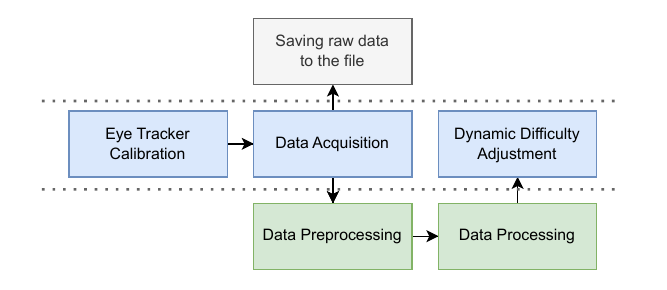}
    \caption{Diagram presenting eye tracking data workflow (source: own elaboration)}
    \label{fig:dataWorkflow}
\end{figure}

Workflow of eye tracking data in our proposition is presented in Figure \ref{fig:dataWorkflow}. As described in subsection \ref{sec:dataAquisition}, first there is eye tracker \textbf{calibration}. After this stage \textbf{data acquisition} is continuous during the XR experience and raw eye tracking data is saved to the file for later analysis.

\textbf{Data preprocessing} focuses on cleaning and refining the raw eye tracking data collected during the XR experience. This process starts with the removal of invalid samples. Next, for each sample, the angular velocity is calculated. And lastly, there is a process of removing one-sample spikes
using Median Filtering\footnote{Median Filtering eliminates outliers by replacing each data point with the median of its neighboring points, which helps smooth out sudden spikes while preserving the overall trend of the data.}.

\textbf{Data processing} involves analyzing the pre-processed eye tracking data to extract meaningful insights into user behavior and cognitive state. This includes detecting \textbf{saccades} and \textbf{fixations}, that are identified using velocity-threshold identification, where rapid shifts in gaze direction exceeding a defined speed indicate a saccade, while fixations are determined by prolonged gaze on a specific point. Additionally, changes in \textbf{pupil diameter} are monitored, serving as an indicator of cognitive load and emotional engagement. Together, these metrics provide valuable information on user attention and mental effort that can be used for \textbf{Dynamic Difficulty Adjustment} in the future adaptive environments made for training, screening, and teleoperation.

\section{Discussion and Further Work}
Having in mind the proposed universal approach towards the eye tracking module, i.e. scalability, reliability, and two-way communication for real-time biofeedback, there are several challenges. One of the most important areas is efficiency, where the challenges relies on the inherent architecture of modern Real-Time Computer Graphics (RTCG) Systems. In particular, the presented module should employ separate threads for data acquisition, preprocessing, and processing to ensure smooth and real-time handling of eye tracking data. This architecture allows for efficient performance, with each stage running independently. For example, Unity’s main thread is closely tied to frame rate (FPS), as it handles rendering, physics, and the core immersive system logic. Overloading the main thread with tasks like data acquisition and processing can lead to frame drops and reduced performance. By offloading these tasks to separate threads, the module will ensure that the main thread remains focused on maintaining smooth flow and high FPS, enhancing overall user experience.

\section*{Acknowledgments}
The authors did not receive any specific grant from funding agencies in the public, commercial, or not-for-profit sectors for this article.

The authors have no competing interests to declare that are relevant to the content of this article.

\bibliographystyle{bibliography/splncs04}
\bibliography{bibliography/bibliography}

\end{document}